\documentclass[11pt]{article}

\usepackage[margin=1in]{geometry}
\usepackage[utf8]{inputenc}
\usepackage[T1]{fontenc}
\usepackage{lmodern}
\usepackage{amsmath}
\usepackage{amsfonts}
\usepackage{amssymb}
\usepackage{dsfont}
\usepackage[normalem]{ulem}
\usepackage{graphicx}
\usepackage{newunicodechar}
\usepackage[affil-it]{authblk}
\usepackage[numbers,sort&compress]{natbib}
\usepackage{hyperref}
\hypersetup{colorlinks=true,citecolor=blue,linkcolor=blue,urlcolor=blue}

\newunicodechar{́}{\'}

\title{Parahydrogen Cooling of Nuclear Spin Chains \\ at Hypogeomagnetic Fields}

\author[1,2]{Alexey Kiryutin}
\author[3]{Danil Markelov}
\author[1,2]{Ivan Zhukov}
\author[4,5,6]{Erik Van Dyke}
\author[1,2]{Alexandra Yurkovskaya}
\author[7,8,9]{Danila Barskiy\thanks{To whom correspondence should be addressed. E-mail: \texttt{barskiy@miami.edu}}}

\affil[1]{International Tomography Center, SB RAS, Novosibirsk, Russia}
\affil[2]{Novosibirsk State University, Novosibirsk, Russia}
\affil[3]{Chimie Physique et Chimie du Vivant (CPCV, UMR 8228), Département de Chimie, École Normale Supérieure, PSL University, Sorbonne Université, Paris 75005, France}
\affil[4]{Johannes Gutenberg University, Institute of Physics, 55128 Mainz, Germany}
\affil[5]{Helmholtz Institute Mainz, 55099 Mainz, Germany}
\affil[6]{GSI Helmholtzzentrum für Schwerionenforschung GmbH, 64291 Darmstadt, Germany}
\affil[7]{Department of Chemistry, University of Miami, Coral Gables, FL 33146, USA}
\affil[8]{Department of Physics, University of Miami, Coral Gables, FL 33146, USA}
\affil[9]{Frost Institute for Chemistry and Molecular Science, Coral Gables, FL 33146, USA}

\date{}

\begin{document}

\maketitle

\begin{abstract}
Solution-state molecular nuclear spin networks are promising quantum simulators because their scalar-coupling Hamiltonians are chemically programmable, precisely measurable, and coherent at room temperature. Their main limitation for quantum information science is initialization: thermal Boltzmann polarization produces highly mixed, high-entropy states. Here,\footnote{This article is based on discussions and experiments conducted before February 2022.} we use parahydrogen-based Signal Amplification by Reversible Exchange (SABRE) at hypogeomagnetic fields (i.e., magnetic fields below Earth field) to hyperpolarize the chemically engineered 12-spin chain [U-$^{13}$C,$^{15}$N]-butyronitrile. SABRE generates percent-level $^{13}$C and $^{15}$N polarization and prepares non-equilibrium multi-spin orders across the network. A von Neumann entropy analysis of such a hyperpolarized system shows that, at the optimal transfer field of 0.52~$\mu$T, the full spin system could reach $S/k_{\mathrm B}=8.274$, compared with $S_{\rm th}/k_{\mathrm B}=8.318$ for the unpolarized reference, giving $(S-S_{\rm th})/k_{\mathrm B}=-0.043$. Experimentally, nuclear spin temperatures of 52\,mK and 257\,mK are achieved for $^{15}$N and $^{13}$C subensembles, respectively. The larger entropy deficit of the full network than of individual subsystems indicates correlated multi-spin order beyond single-spin polarizations. Rapid field cycling to 9.4~T enables site-resolved NMR readout, while the precisely determined coupling network provides an experimentally benchmarked Hamiltonian for testing quantum-simulation, quantum-control, and Hamiltonian-learning protocols.
\end{abstract}


\\

\section*{Introduction}

Quantum computation promises to outperform classical approaches in a variety of tasks ranging from integer factorization in cryptography \cite{portmann2022security, bennett1992experimental} to the simulation of complex molecular and quantum many-body systems \cite{daley2022practical, divincenzo2000physical, nielsen2010quantum, simon1997power}. Among the physical platforms explored for implementing quantum information processing, nuclear magnetic resonance (NMR) has historically played an important role as a testbed for small-scale quantum algorithms \cite{jones1998implementation_alg, glaser2001nmr, linden1998implementation, chuang1998experimental, cory1997ensemble}. Early demonstrations, including implementations of Shor's factoring algorithm \cite{vandersypen2001experimental} and Grover's search \cite{jones1998implementation_seacrh}, showed that coupled networks of nuclear spins can serve as controllable quantum registers, where radiofrequency pulse sequences implement unitary operations and NMR spectra provide direct readout of statistically averaged quantum states \cite{gershenfeld1997bulk}. Carefully designed pulse sequences allow the engineering of effective Hamiltonians and high-fidelity quantum gates \cite{o2019hamiltonian, zhou2023robust, zhou2024robust, choi2020robust, geier2021floquet, ajoy2012algorithmic, ajoy2013quantum, alway2007arbitrary}, which have enabled a wide range of proof-of-principle experiments demonstrating coherent control of multi-spin systems \cite{marx2000approaching, vandersypen2004nmr}.

Individual nuclear spins are natural qubits, with scalar $J$-couplings mediating interactions necessary for multi-qubit operations. These couplings also facilitate the propagation of quantum correlations and the transfer of spin order along the network \cite{sorensen1984product,kiryutin2026high}. Nuclear spins are particularly attractive for such studies because their $J$-coupling constants are well characterized and coherence times can remain long ($T^{*}_2 \gg 1/J$)
even at room temperature. One particularly interesting realization of nuclear spin networks are one-dimensional spin chains, which, in practice, can be realized by using molecules in solution \cite{kiryutin2026high, sheberstov2024collective, sonnefeld2022long, sonnefeld2022polychromatic, sheberstov2025aliphatic}. They provide controllable and chemically tunable platforms for exploring quantum transport, entanglement growth, and many-body spin dynamics in chemically and biologically relevant environments. 

While molecular spin chains offer a versatile framework for studying quantum dynamics, a fundamental limitation of conventional liquid-state NMR quantum computing arises: thermal equilibrium polarization of nuclear spins is extremely small, typically not exceeding 0.001--0.01\%. Therefore, the initial state of such spin ensemble is highly mixed, and most implementations to date have relied on the so-called pseudo-pure states that emulate pure-state quantum computation within a largely unpolarized spin ensemble \cite{chuang1998bulk,peng2001preparation, sharf2000spatially, pravia1999observations, gong2024complexity}. Although this approach has enabled pioneering demonstrations of quantum algorithms, the exponentially decreasing purity with system size severely limits scalability \cite{warren_science}.

Hyperpolarization methods offer a promising route for overcoming intrinsically low thermal polarization in NMR.
Several techniques enable nuclear spin polarization to $\gtrsim$$10\%$ \cite{ardenkjaer2003increase, rayner2017delivering}, therefore, potentially providing the critical state-preparation resource to increase the initial state purity often required for quantum computing and quantum simulations \cite{warren_science, berry2025rapid, weber2022toward,rosenkranz2025quantum}. In particular, the enhanced polarization allows the preparation of highly ordered initial states (pre-initialization) of interacting nuclear spins.
The degree of such spin ordering can be quantified thermodynamically through the spin system entropy: lower value corresponds to a more ordered nuclear spin ensemble. For Zeeman-polarized spin subsystems, the same non-equilibrium order can also be expressed as an effective spin temperature, providing an intuitive metric for how far the nuclear spin populations have been driven from room-temperature Boltzmann equilibrium. Importantly, pre-initialization does not need to cool the sample itself, but it should reduce the entropy and effective spin temperature of the nuclear spin degrees of freedom.
Although the ensembles can remain mixed, the substantially increased polarization and the presence of multispin orders are expected to be readily employed for preparing pseudo-pure states via pulse sequences/adiabatic protocols, facilitating the development of
experiments with nuclear-based multispin registers applicable toward a wider range of problems compared to previously conducted liquid-state-NMR-based demonstrations \cite{du2010nmr}.

\begin{figure}[!htb]

    \centering
    \includegraphics[width=11.4cm]{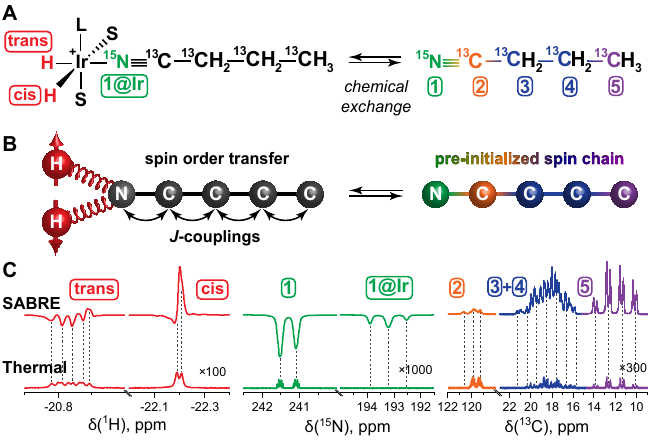}
    \caption{Signal Amplification by Reversible Exchange (SABRE) as a method for pre-initializing nuclear spin chains. (A)~Schematic of the reversible exchange of parahydrogen and a [U-$^{13}$C,$^{15}$N]-butyronitrile spin chain with an Ir-based polarization transfer complex. L and S denote ancillary ligands of the complex. (B) Polarization transfer from parahydrogen to the spin chain occurs through the network of scalar \textit{J}-couplings within the complex, resulting in hyperpolarization of the spin chain. (C) $^{1}$H, $^{15}$N, and $^{13}$C NMR signals hyperpolarized by SABRE at 1 $\mu$T (top traces) compared with the corresponding thermal spectra, scaled by factors of 100, 1000, and 300, respectively (bottom traces). Note that some signals (e.g., 1@Ir) in the top SABRE-polarized spectra correspond to the Ir-complex-bound spin chain molecules.} \label{fig1}
\end{figure}

Among hyperpolarization strategies, parahydrogen provides a particularly attractive source of nuclear spin order for quantum initialization \cite{bowers1987parahydrogen, anwar2004implementing, blazina2005generation, hubler2000nuclear}. Parahydrogen ($p$H$_2$) is the singlet nuclear-spin isomer of the $\text{H}_2$ molecule, prepared by cooling the gas to cryogenic temperatures in the presence of paramagnetic catalysts ($\sim$25\,K). Although NMR-silent, this singlet spin order can be converted into observable nuclear polarization in the products of $p$H$_2$ addition reactions. The resulting polarization corresponds to an effective spin temperature of as low as 6.4~mK, making parahydrogen-derived spin order a highly non-equilibrium resource for quantum information processing \cite{anwar2004preparing}.
Although the idea of using $p$H$_2$ as a source of quantum state initialization is not new, it has only been demonstrated on a two-atom system of a metal hydride without subsequent polarization transfer. In this work, we extend the scope of quantum pre-initialization from two to 12 spins by using an exemplary nuclear spin chain system with a well defined coupling network \cite{kiryutin2026high}.

Signal Amplification by Reversible Exchange (SABRE) is a $p$H$_2$-based hyperpolarization technique that has emerged as a particularly versatile approach for generating substantial nuclear polarization \cite{adams2009reversible, rayner2018signal}.
In SABRE, spin order from $p$H$_2$ is transferred to the target nuclei via reversible exchange with a metal complex, producing hyperpolarization without modifying the substrate chemically; because the exchange cycle is reversible, hyperpolarization can be regenerated continuously as long as fresh $p$H$_2$ is supplied. Under hypogeomagnetic (microtesla) magnetic fields, strong mixing of spin states allows efficient propagation of polarization from $p$H$_2$ to heteronuclei through the network of scalar couplings, distributing it across multiple spins \cite{theis2015microtesla,truong201515n}. The coherent nature of this transfer has been directly observed via oscillatory polarization dynamics when passing through level anti-crossings (LACs) in model three-spin systems \cite{kiryutin2013evidence}, confirming that interactions with $p$H$_2$ can establish correlated multispin orders rather than merely independent polarizations. This ability to channel polarization along coupled-spin networks as well as ease with which multiple experiments can be repeated on the same sample, makes SABRE especially well suited for preparing highly spin-ordered nuclear spin chains at room temperature. Such spin chains are poised to become fruitful platforms for quantum information simulation/processing experiments as well as artificial intelligence-enhanced quantum simulation approaches \cite{alexeev2025artificial}.

In this work, we employ SABRE to hyperpolarize nuclear spin chain [U-$^{13}$C,$^{15}$N]-butyronitrile, generating enhanced and distributed polarization across a network of coupled nuclear spins. This effectively corresponds to spin cooling down to $\sim$50\,mK, albeit the chemical system remains at steady state and room temperature. The capacity of SABRE to hyperpolarize multiple spins simultaneously provides a route to preparing spin ensembles for quantum information processing and quantum simulation in liquid-state NMR. Rather than implementing a specific algorithm, the goal of this work is to establish a valid method for pre-initializing (beyond Boltzmann polarization) multi-spin registers that can serve as platforms for quantum control and computation experiments. SABRE-generated polarization produces pre-initialization, i.e., multi-spin states with substantially higher purity than conventional thermal equilibrium states, bringing them closer to the ideal initial states required in quantum information processing. Our approach opens the doors to exploring larger highly polarized spin networks, expanding the experimental capabilities of NMR-based quantum information science.

\section*{Results}

Figure~1 schematically introduces the SABRE-based pre-initialization scheme used in this work. Reversible exchange of $p$H$_2$ and [U-$^{13}$C,$^{15}$N]-butyronitrile on the Ir-based polarization-transfer complex (with L denoting IMes = 1,3-bis(2,4,6-trimethylphenyl)-1,3-dihydro-2H-imidazol-2-ylidene, and S denoting pyridine as the ancillary ligand) together with coherent transfer of spin order from the ligated $p$H$_2$ atoms results in the accumulation of polarization on the free spin-chain molecules (Fig.~1A). Through the $J$-coupling network, this process prepares ordered states of the coupled nuclear-spin ensemble in the bulk sense \cite{chuang1998bulk}. Importantly, SABRE polarizes all nuclei in the molecule simultaneously \cite{kiryutin2016fast}, and the specific spin orders that are created depend on the magnetic field at which polarization transfer takes place as well as on the field manipulations applied during the experiment (Fig.~1B) \cite{eriksson2022,lindale2024,kozinenko2026}.

Representative SABRE-polarized and thermally polarized $^1$H (hydride region), $^{15}$N, and $^{13}$C NMR spectra are shown in Fig.~1C. The upper traces correspond to SABRE-enhanced signals obtained at a transfer field of 1~$\mu$T, whereas the lower traces show the corresponding thermal spectra scaled by factors of 100, 1000, and 300, respectively. After only a few seconds of $p$H$_2$ bubbling through the solution, SABRE yields signal enhancements ranging from several hundred to a few thousand. In addition to signals from free [U-$^{13}$C,$^{15}$N]-butyronitrile, the SABRE spectra also contain contributions from bound spin-chain molecules: for example, ``1@Ir'' denotes the nitrile $^{15}$N atom bound to the polarization-transfer complex, and additional bound-state peaks are also observed for some carbon atoms (e.g., for C$_2$).

The line shapes in the $^{15}$N NMR spectra reflect the coupling topology of the spin system. The $^{15}$N signal of free [U-$^{13}$C,$^{15}$N]-butyronitrile appears as a doublet because the nitrile $^{15}$N is split by the directly bonded $^{13}$C (17.37\,Hz). By contrast, the corresponding signal of the bound spin-chain molecule (1@Ir) appears as a triplet. This pattern arises from the combined action of the $^{15}$N-$^{13}$C coupling within the nitrile group and the coupling to the trans-hydride in the Ir complex (see Supporting Information, SI, for details). Because the $^{15}$N-$^1$H and $^{15}$N-$^{13}$C couplings are of similar magnitude ($J_{\rm NH} = -25.6$\,Hz, $J_{\rm NC} = 30.3$\,Hz), the resulting multiplet appears triplet-like.

Figure~2 compares SABRE-enhanced $^{15}$N and $^{13}$C NMR spectra recorded after $p$H$_2$ bubbling at ultralow magnetic fields, between --0.5~$\mu$T and +0.9~$\mu$T, where the sign indicates the field direction relative to the detection field of the high-field NMR spectrometer. For both types of nuclei, $^{15}$N and $^{13}$C, the spectra display strongly enhanced lines with both positive and negative contributions, indicating the formation of various multispin orders in the chain molecule.

\begin{figure}[!htb]

    \centering
    \includegraphics[width=8.7cm]{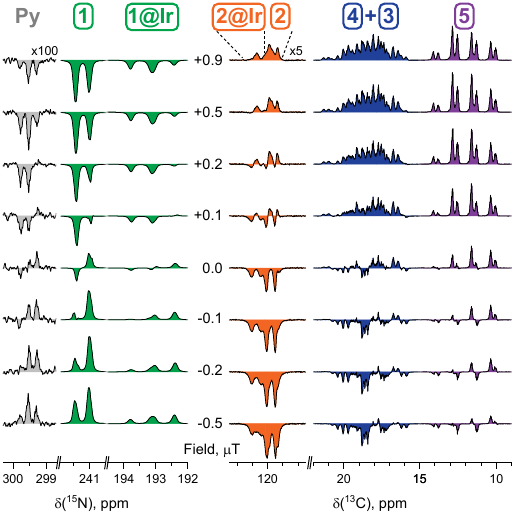}
    \caption{$^{15}$N (left) and $^{13}$C (right) NMR spectra (9.4~T) of [U-$^{13}$C,$^{15}$N]-butyronitrile obtained after SABRE polarization at ultralow magnetic fields (from $-0.5~\mu$T to $+0.9~\mu$T). The negative sign of the magnetic field indicates the opposite direction relative to the detection field in the NMR magnet. Note the multiplication factors of $\times100$ and $\times5$ for the data columns corresponding to pyridine (Py) and 2@Ir/2, respectively.}

    \label{fig2}
\end{figure}

The intensities of the individual $^{15}$N lines strongly depend on the polarization-transfer field. At zero magnetic field, $^{15}$N NMR spectra consist of antiphase multiplets, such that the total integral of each multiplet is close to zero. The same qualitative behavior is observed for pyridine (Fig.~2, left), which serves as a co-ligand to stabilize metal complex (Ir complex with only butyronitrile ligands is unstable). $^{15}$N NMR signal of pyridine (nat. abun.) appears as a triplet because of splitting by the two equivalent ortho-protons in the aromatic ring. Overall, the sign of the $^{15}$N polarization changes approximately symmetrically around zero field.

The $^{13}$C NMR spectra behave differently. Unlike the response of the $^{15}$N spins, the $^{13}$C spectra display a pronounced asymmetry with respect to the sign of the applied transfer field. This asymmetry is not a property of our SABRE step; rather, it reflects polarization/spin order redistribution among the carbon spins during sample transfer from the field at which $p$H$_2$ is introduced into the system to the high-field NMR spectrometer. Although both coherent transfer through the $J$-coupling network and incoherent cross-relaxation may occur, the dominant contribution is expected to be coherent \cite{kiryutin2026high}.
It is during this transport stage ($\approx$400 ms) that redistribution of non-equilibrium polarizations occurs among spins of the same type (e.g., C$_2$-C$_5$) due to the strong coupling in field regimes ranging from $\mu$T to mT. This effect is analogous to the well known ALTADENA experiment \cite{pravica1988net,kiryutin2026high}, in which the two outermost NMR lines of a coupled spin system acquire strongly enhanced polarizations of opposite sign, so that one end of the spectrum appears positive and the other negative \cite{Plaumann2013,Kiryutin2024Photo}.
The contrast between pyridine and the spin chain shows that the dynamics of the spin chain molecule are exceptionally sensitive to tiny magnetic fields, those that even NMR features of SABRE-polarized [$^{15}$N]-pyridine cannot capture.

Figure 3A directly summarizes the field dependence of SABRE polarization in the ultralow-field range from $-4$\,$\mu$T to $+4$\,$\mu$T, where states of different spin types ($^{15}$N, $^{13}$C, and $^{1}$H) mix. In the upper part of the panel, the experimental signal intensities obtained after 10\,s of $p$H$_2$ bubbling at 25\,$^\circ$C are plotted as a function of the transfer field. The intensities are reported as percentages of the maximum theoretical polarization. For the $^{15}$N spin, the field dependence is an odd function of the magnetic field, implying that the integrated signal of 1 and 1@Ir vanishes at zero field, and the polarization reaches a maximum at about $2$\,$\mu$T. The $^{13}$C spins also show a maximum, but at a lower field of about $0.5\,\mu$T.

The lower part of Fig.~3A shows the corresponding theoretical SABRE calculation, which includes all spins but assumes instantaneous field switching. For the $^{15}$N spin, the calculation reproduces the antisymmetry about zero field very well and predicts the polarization maximum at $1$--$2\,\mu$T. For the carbon spins, it captures the overall field dependence and the position of the SABRE maximum reasonably well, but clear discrepancies remain. In the calculation, C$_2$ has the highest polarization among the $^{13}$C nuclei, whereas experimentally it has the lowest. In contrast, the distant chain spin C$_5$ is only weakly polarized in the calculation but shows the strongest experimental enhancement, consistent with the spectra in Fig.~2 (note the multiplication factor of $5$ for all ``2/2@Ir'' NMR spectra). The calculation also fails to reproduce the asymmetry of the $^{13}$C signals about zero field. Both discrepancies likely arise from the experimental field-switching protocol: after a rapid electromagnetic switch to $50$\,$\mu$T, the sample is moved out of the magnetic shield, and during this slower second stage the field increases smoothly from $50$\,$\mu$T to the detection field of 9.4\,T. During this $\approx$400\,ms interval, coherent redistribution of non-equilibrium polarization can occur among spins of the same type (e.g., C$_2$--C$_5$) because of strong coupling in the $\mu$T-to-mT range, altering the observable signal intensities compared to those expected from the sudden-switch experiments.

Figure 3B shows the buildup of SABRE polarization in [U-$^{13}$C,$^{15}$N]-butyronitrile at two representative transfer fields, 0.4 and 2.0\,$\mu$T. The thin black lines are mono-exponential fits used to extract the effective buildup time constant, $\tau_{\mathrm{eff}}$. The growth of $^{15}$N polarization is well described by a mono-exponential function, yielding time constants of 2.5\,s and 3.8\,s at 0.4\,$\mu$T and 2.0\,$\mu$T, respectively. The buildup is much faster than the high-field relaxation times of $^{15}$N and $^{13}$C and is therefore expected to be governed by the collective relaxation dynamics of the coupled butyronitrile spin system while the molecule is bound to the iridium complex. This field dependence of the buildup time should be carefully considered in other SABRE experiments that map polarization as a function of the transfer field, for example when $p$H$_2$ is bubbled continuously while the magnetic field is varied. If a single buildup time optimized for one field is used across the entire field profile, polarization at other fields may not have reached its plateau, leading to an apparent field dependence that reflects incomplete buildup rather than the true SABRE transfer efficiency at a given field.

The lower panel of Fig.~3B shows the corresponding buildup of $^{13}$C polarization. In contrast to the $^{15}$N case, the $^{13}$C buildup at 2.0\,$\mu$T exhibits a pronounced slow initial phase, indicating that the polarization dynamics are not purely mono-exponential but are shaped by subsequent redistribution through the $J$-coupling network. Together, Figs.~3A and 3B provide a compact description of the field dependence and buildup dynamics of SABRE polarization in this spin-chain system.

\section*{Discussion}

\begin{figure}

    \centering 
    \includegraphics[width=11.4cm]{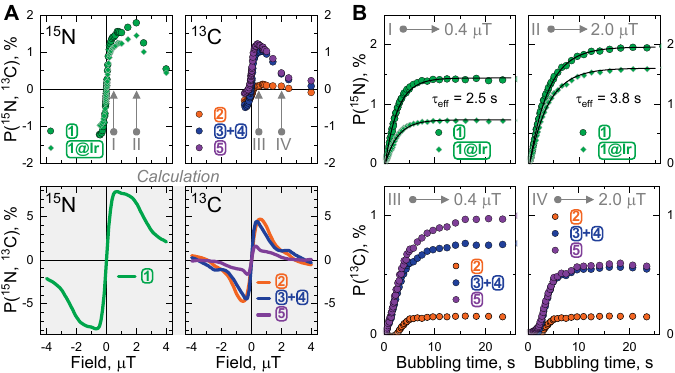}
    \caption{(A) Experimental (top) and calculated (bottom) dependence of signal intensities of SABRE-polarized signals as a function of ultralow magnetic field during parahydrogen bubbling. The bubbling time was set to 10 s for all experiments, and the temperature was maintained at 25~$^\circ$C; (B) Dynamics of SABRE polarization buildup measured at magnetic fields of 0.4~$\mu$T and 2.0~$\mu$T for the $^{15}$N (top) and $^{13}$C (bottom) spins in [U-$^{13}$C, $^{15}$N]-butyronitrile. All signal intensities are given as a percentage of the maximum theoretical polarization. The thin black lines show fits to the data using a mono-exponential function with the effective buildup time constant, $\tau_{\mathrm{eff}}$.}
    \label{fig3}
\end{figure}

We calculated the von Neumann entropy, $S= \nobreak -k_{\mathrm B} \operatorname{Tr}(\hat{\rho} \ln \hat{\rho})$, for the nuclear spin
chain across a range of magnetic fields at which SABRE polarization transfer takes place (here, $\hat{\rho}$ is a density matrix of [U-$^{13}$C,$^{15}$N]-butyronitrile  and $k_{\mathrm B}$ is the Boltzmann's constant). The reference state was taken as the completely unpolarized ensemble, neglecting the thermal polarization at room temperature \cite{levitt2021hyperpolarization}. The entropy was evaluated for the four cases: the full spin system (all $^{1}$H, $^{13}$C, and $^{15}$N nuclei), the heteronuclear subsystem ($^{13}$C and $^{15}$N only), the carbon subsystem ($^{13}$C spins only), and the nitrogen subsystem ($^{15}$N only), as shown in Table~\ref{tab:entropy}. The largest deviation of the entropy from thermal equilibrium, $(S - S_{\rm th})/k_{\mathrm B} = -4.3 \times 10^{-2}$, was observed for the whole system at the magnetic field of $0.52$\,$\mu$T, which coincides with the numerically calculated maximum of $^{15}$N and $^{13}$C polarization, as shown in Fig.~\ref{fig3}A. Notably, this deviation is greater than that of any individual spin subsystem, indicating the efficient generation of not only polarization but also of multi-spin correlations among $^{15}$N, $^{13}$C, and $^{1}$H nuclei (see Fig.\,S1). This result demonstrates that SABRE methodology significantly reduces the entropy of the spin chain, driving the system toward a more ordered state, pre-initialized for subsequent quantum information protocols. For reference, thermal polarization of the same spin system at 9.4 T reduces the entropy relative to the fully unpolarized state by merely $10^{-9}$. The SABRE-induced entropy reduction therefore exceeds this value by more than seven orders of magnitude, demonstrating the remarkable capability of SABRE to create highly ordered spin states, in a few seconds and at room temperature.

\begin{table}[ht]
\centering
\caption{Deviations of  [$^{13}$C,$^{15}$N]-butyronitrile's spin entropy from thermal equilibrium and effective spin temperatures at 10\,T corresponding to the SABRE pre-initialization field of 0.52\,$\mu$T. \\ }
\label{tab:entropy}
\begin{tabular}{lcccc}
\hline
\textbf{Spin system} & \textbf{$S/k_{\mathrm B}$} & \textbf{$S_{\rm th}/k_{\mathrm B}$} & \textbf{$(S$$-$$S_{\rm th})/k_{\mathrm B}$\,$\times$\,$10^{2}$} & $T_{\mathrm E}$,\,mK \\
\hline
Full (12 spins) & $8.274$ & $8.318$ & $-4.3$ & $-$ \\
$^{15}$N, $^{13}$C (5 spins) & $3.438$ & $3.466$ & $-2.7$ & $-$ \\
$^{15}$N (1 spin) & $0.680$ & $0.693$ & $-1.3$ & $52$ \\
$^{13}$C (4 spins) & $2.762$ & $2.773$ & $-1.1$ & $257$ \\
\hline
\end{tabular}
\vspace{2pt}
\begin{minipage}{\linewidth}
\footnotesize
$k_{\mathrm B} = 1.38 \times 10^{-23}$ [J/K] is the Boltzmann constant;\\
$S_{\text{th}} = N k_{\mathrm B} \ln 2$ is the entropy for $N$ unpolarized spin-1/2 nuclei; $T_{\mathrm E}$ is experimentally measured spin temperature.
\end{minipage}
\end{table}

If all available spin orders were to transfer into a single transition, i.e., corresponding to $^{15}$N spin flips, S{\o}rensen bounds can give an estimate \cite{SORENSEN1990435}. We concluded that a pulse sequence/adiabatic protocol with rational design could therefore improve the achieved signal intensities and also produce an appropriate initial state with a substantial purity of $\sim$15\,\% (see SI).

Because the spin system evolves under a fully characterized spin-coupling Hamiltonian, the experiment directly implements analog quantum simulation of a Heisenberg spin chain: measured time-domain observables provide ground-truth dynamics that can be compared against classical calculations and, in future work, against quantum algorithms attempting to reproduce the same evolution under experimentally realistic control constraints.

More broadly, the combination of zero- to ultralow-field (ZULF) evolution with high-field detection cleanly separates Hamiltonian dynamics from site-resolved readout. In the ZULF regime, chemical-shift dispersion is quenched and the isotropic scalar-coupling Hamiltonian is not truncated by dominant Zeeman terms, enabling strong heteronuclear couplings and Heisenberg-type dynamics. High-field readout then restores both sensitivity and chemical-shift resolution, allowing site-specific spectral assignment. Magnetic field cycling therefore provides a practical interface between an analog many-body simulator and a conventional high-field NMR measurement layer.

The same pre-initialization can also be expressed in terms of an effective Zeeman spin temperature, $T_{\mathrm E}$. For a spin-$1/2$ nucleus with polarization $P$ in a magnetic field of magnitude $B$, inversion of the Boltzmann relation gives $T_{\mathrm E} = |\hbar \gamma B/(2k_{\mathrm B}\operatorname{arctanh}(P))|$.
Using $|\gamma_{^{15}\mathrm N}|/2\pi = 4.316$~MHz~T$^{-1}$, a $^{15}$N polarization of 20\% at $\simeq$10~T corresponds to $\simeq$5~mK. At the SABRE transfer field, the same thermometric scale becomes dramatically more stringent: achieving even 1\% thermal $^{15}$N polarization at $0.5~\mu$T would require $\simeq$5~nK. For comparison, the lowest artificial matter temperature reported to date, 38~pK \cite{deppner2021collectivemode}, would correspond to $P_{^{15}\mathrm N}\simeq88$\,\% at $0.5~\mu$T. These estimates emphasize that SABRE does not cool the liquid sample itself; rather, it prepares a room-temperature ensemble whose nuclear Zeeman populations, multispin correlations, and von Neumann entropy correspond to a spin reservoir far outside the range accessible by ordinary Boltzmann polarization. To our knowledge, such a quantitative entropy and spin-temperature assignment for SABRE-pre-initialized, liquid-state molecular spin-chain registers has not previously been reported.

A natural next step is to convert SABRE-prepared, lower-entropy spin ensembles into targeted initial states for quantum-simulation and quantum-control experiments. The present platform already provides a useful set of native operations: magnetic-field quenches between weak- and strong-coupling regimes, evolution under the intrinsic isotropic $J$-coupling Hamiltonian at ZULF, global DC control at ultralow field, selective RF control and site-resolved detection at high field, and hyperpolarized pre-initialization from $p$H$_2$. Together with an independently benchmarked coupling network, these capabilities make the system well suited for algorithm--hardware co-design, where pulse sequences, field-cycling trajectories, and state-preparation protocols are optimized for the experimentally available operations rather than imposed from an idealized gate model.

Such calibrated molecular spin chains can therefore serve as compact, reproducible testbeds for near-term quantum science. They provide experimentally accessible examples in which the Hamiltonian is known, the initial state can be driven far beyond thermal Boltzmann polarization, and the resulting dynamics can be measured with high spectral resolution.
This combination is valuable for validating Hamiltonian-learning methods \cite{schwade2026physics}, benchmarking time-evolution algorithms \cite{lin2021real,Yeter2021}, testing strategies for generating and detecting multispin correlations \cite{ALBERT2017293,CADARS200724}, and exploring how non-equilibrium spin order propagates through chemically defined networks \cite{ueda2023non}. Because the molecular scaffold can be varied synthetically, the same approach can be extended to spin chains with different lengths, topologies, coupling strengths, and relaxation environments.

Overall, this work establishes parahydrogen-driven SABRE hyperpolarization at hypogeomagnetic (i.e., ZULF) fields as a practical route for pre-initializing molecular nuclear spin chains. In [U-$^{13}$C,$^{15}$N]-butyronitrile, SABRE produces percent-level heteronuclear polarization, creates correlated multispin order, and measurably reduces the entropy of a 12-spin network within seconds at room temperature. By combining this beyond-Boltzmann initialization with field-cycling control and high-field NMR readout, the approach bridges hyperpolarization chemistry, precision spectroscopy, and quantum simulation. These developments point toward increasingly capable liquid-state NMR platforms in which chemically programmable molecules act as calibrated, hyperpolarized quantum simulators for studying coherent spin dynamics in realistic many-spin systems.

\section*{Materials and Methods}

\subsection*{NMR Experiments}

NMR experiments were performed using a home-built mechanical field-cycling setup based on a 400 MHz NMR spectrometer, equipped with a magnetic field shield installed on top of the cryomagnet \cite{zhukov2018field}. This versatile setup enables both magnetic field cycling and in-situ sample bubbling with parahydrogen under elevated pressure across a wide range of magnetic fields \cite{kiryutin2018sabre}. High-resolution NMR spectra for spin system simulation were acquired on a separate NMR spectrometer with a base proton frequency of 700 MHz at ambient temperature.

As a source of non-thermal polarization, we utilized parahydrogen gas (85\% enrichment, Bruker parahydrogen generator BPHG-90) in the Signal Amplification By Reversible Exchange (SABRE) process. The solution for SABRE experiments contained 12.9 mM of [U-$^{13}$C,$^{15}$N]-butyronitrile, 20 mM of $^{14}$N-pyridine as a co-ligand, and 5.3 mM of the precatalyst [IrCl(COD)(IMes)] in deuterated methanol-$d_4$. To activate the precatalyst by hydrogenation of cyclooctadiene (COD) and form the stable iridium complex with pyridine and butyronitrile, the solution was purged with parahydrogen gas at 3 bar for approximately 15 minutes. SABRE experiments were performed at 20~$^\circ$C with a bubbling time of 10 seconds using the aforementioned field-cycling apparatus. The setup includes an add-on for in-situ bubbling in a standard NMR sample tube at elevated pressures up to 10 bars; further details of this apparatus can be found in references \cite{kiryutin2018sabre, kiryutin2019phipsetup}.

\subsection*{$J$-coupling determination by fitting of NMR spectra}
$^1$H, $^{13}$C, and $^{15}$N high resolution NMR spectra were fitted by ANATOLIA software \cite{ANATOLIA}. The protons in the CH$_3$-CH$_2$-CH$_2$- group were treated as an A$_3$M$_2$X$_2$ system; since the protons in CH$_2$ and CH$_3$ groups are magnetically equivalent, the coupling within the groups does not influence NMR spectra. The errors in the determinations of the spin-spin couplings are in a range of 0.01-0.05 Hz.

\subsection*{Chemicals}
[U-$^{13}$C, $^{15}$N]-butyronitrile was obtained from Sigma-Aldrich via custom synthesis, standard iridium precatalyst for SABRE  [IrCl(COD)(IMes)] was synthetized as described in \cite{kownacki2008synthesis} . Methanol-$d_4$ (99.8\%) was purchased from Carl-Roth (Germany). $^{14}$N-pyridine (99.8\%) was purchased from Sigma Aldrich.

\subsection*{Supporting Information}

Details of the numerical simulations, NMR parameters and spectra of bound [U-$^{13}$C,$^{15}$N]-butyronitrile, as well as experimental details for SABRE generation at hypogeomagnetic fields are provided in the Supplementary Information (SI).

\section*{Acknowledgments}
A.K., D.M., I.Z., and A.Y. acknowledges the Ministry of Science and Higher Education of the Russian Federation for granting
access to the equipment. D.B. and E.V.D. acknowledge the Alexander von Humboldt Foundation for the financial support provided in the framework of the Sofja Kovalevskaja Award, endowed by the German Federal Ministry of Education and Research.

\section*{Author Contributions}
A.K.: data acquisition, methodology, investigation, visualization, writing original draft, review \& editing;
D.M.: investigation, numerical simulations, visualization, writing original draft, review \& editing;
I.Z.: investigation, visualization, writing original draft, review \& editing; E.V.D.: investigation, review \& editing; A.Yu.: review \& editing, supervision, resources;
D.B.: conceptualization, methodology, investigation, writing original draft, visualization, review \& editing, project administration.

\section*{Competing Interests}
The authors declare no competing interests.

\bibliographystyle{unsrt}
\bibliography{references}

@article{warren_science,
author = {Warren S. Warren },
title = {The Usefulness of {NMR} Quantum Computing},
journal = {Science},
volume = {277},
number = {5332},
pages = {1688-1690},
year = {1997},
doi = {10.1126/science.277.5332.1688},
URL = {https://www.science.org/doi/abs/10.1126/science.277.5332.1688},
eprint = {https://www.science.org/doi/pdf/10.1126/science.277.5332.1688}}

@article{rayner2017delivering,
  title={Delivering strong 1{H} nuclear hyperpolarization levels and long magnetic lifetimes through signal amplification by reversible exchange},
  author={Rayner, Peter J and Burns, Michael J and Olaru, Alexandra M and Norcott, Philip and Fekete, Marianna and Green, Gary GR and Highton, Louise AR and Mewis, Ryan E and Duckett, Simon B},
  journal={Proc. Nat. Acad. Sci.},
  volume={114},
  number={16},
  pages={E3188--E3194},
  year={2017},
  publisher={National Acad Sciences}
}

@article{kiryutin2016fast,
  title={A fast field-cycling device for high-resolution {NMR}: {D}esign and application to spin relaxation and hyperpolarization experiments},
  author={Kiryutin, Alexey S and Pravdivtsev, Andrey N and Ivanov, Konstantin L and Grishin, Yuri A and Vieth, Hans-Martin and Yurkovskaya, Alexandra V},
  journal={J. Magn. Reson.},
  volume={263},
  pages={79--91},
  year={2016},
  publisher={Elsevier}
}

@article{ANATOLIA,
author = {Cheshkov, D.A. and Sheberstov, K.F. and Sinitsyn, D.O. and Chertkov, V.A.},
title = {{ANATOLIA}: {NMR} software for spectral analysis of total lineshape},
journal = {Magnetic Resonance in Chemistry},
volume = {56},
number = {6},
pages = {449-457},
doi = {https://doi.org/10.1002/mrc.4689},
url = {https://analyticalsciencejournals.onlinelibrary.wiley.com/doi/abs/10.1002/mrc.4689},
eprint = {https://analyticalsciencejournals.onlinelibrary.wiley.com/doi/pdf/10.1002/mrc.4689},
year = {2018}
}

@article{zhukov2018field,
  title={Field-cycling {NMR} experiments in an ultra-wide magnetic field range: relaxation and coherent polarization transfer},
  author={Zhukov, Ivan V and Kiryutin, Alexey S and Yurkovskaya, Alexandra V and Grishin, Yuri A and Vieth, Hans-Martin and Ivanov, Konstantin L},
  journal={Phys. Chem. Chem. Phys.},
  volume={20},
  number={18},
  pages={12396--12405},
  year={2018},
  publisher={Royal Society of Chemistry}
}

@article{kiryutin2019phipsetup,
  title={A highly versatile automatized setup for quantitative measurements of {PHIP} enhancements},
  author={Kiryutin, Alexey S. and Sauer, Grit and Hadjiali, Sara and Yurkovskaya, Alexandra V. and Breitzke, Hergen and Buntkowsky, Gerd},
  journal={J. Magn. Reson.},
  volume={285},
  number={18},
  pages={26--36},
  year={2017},
  publisher={Elsevier}
}

@article{kiryutin2018sabre,
  title={Complete magnetic field dependence of {SABRE}-derived polarization },
  author={Kiryutin, Alexey S. and Yurkovskaya, Alexandra and Zimmerman, Herbert and Vieth, Hans-Martin and Ivanov, Konstantin L.},
  journal={Magn. Reson. Chem},
  volume={56},
  number={7},
  pages={651--662},
  year={2018},
  publisher={Wiley Online Library}
}

@article{kownacki2008synthesis,
  title={Synthesis, structure and catalytic activity of the first iridium ({I}) siloxide versus chloride complexes with 1, 3-mesitylimidazolin-2-ylidene ligand},
  author={Kownacki, Ireneusz and Kubicki, Maciej and Szubert, Karol and Marciniec, Bogdan},
  journal={J. of Organomet. Chem.},
  volume={693},
  number={2},
  pages={321--328},
  year={2008},
  publisher={Elsevier}
}

@book{nielsen2010quantum,
  title={Quantum computation and quantum information},
  author={Nielsen, Michael A and Chuang, Isaac L},
  year={2010},
  publisher={Cambridge university press}
}

@article{simon1997power,
  title={On the power of quantum computation},
  author={Simon, Daniel R},
  journal={SIAM J. Comput.},
  volume={26},
  number={5},
  pages={1474--1483},
  year={1997},
  publisher={SIAM}
}

@article{divincenzo2000physical,
  title={The physical implementation of quantum computation},
  author={DiVincenzo, David P},
  journal={Fortschr. Phys.},
  volume={48},
  number={9-11},
  pages={771--783},
  year={2000},
  publisher={Wiley Online Library}
}

@article{jones1998implementation_alg,
  title={Implementation of a quantum algorithm on a nuclear magnetic resonance quantum computer},
  author={Jones, Jonathan A and Mosca, Michele},
  journal={J. Chem. Phys.},
  volume={109},
  number={5},
  pages={1648--1653},
  year={1998},
  publisher={American Institute of Physics}
}

@article{glaser2001nmr,
  title={{NMR} quantum computing},
  author={Glaser, Steffen J},
  journal={Angew. Chem. Int. Ed.},
  volume={40},
  number={1},
  pages={147--149},
  year={2001},
  publisher={Wiley Online Library}
}

@article{linden1998implementation,
  title={An implementation of the {D}eutsch--{J}ozsa algorithm on a three-qubit NMR quantum computer},
  author={Linden, Noah and Barjat, Herv{e} and Freeman, Ray},
  journal={Chem. Phys. Lett.},
  volume={296},
  number={1-2},
  pages={61--67},
  year={1998},
  publisher={Elsevier}
}

@article{vandersypen2001experimental,
  title={Experimental realization of {S}hor's quantum factoring algorithm using nuclear magnetic resonance},
  author={Vandersypen, Lieven MK and Steffen, Matthias and Breyta, Gregory and Yannoni, Costantino S and Sherwood, Mark H and Chuang, Isaac L},
  journal={Nature},
  volume={414},
  number={6866},
  pages={883--887},
  year={2001},
  publisher={Nature Publishing Group UK London}
}

@article{alexeev2025artificial,
  title={Artificial intelligence for quantum computing},
  author={Alexeev, Yuri and Farag, Marwa H and Patti, Taylor L and Wolf, Mark E and Ares, Natalia and Aspuru-Guzik, Alan and Benjamin, Simon C and Cai, Zhenyu and Cao, Shuxiang and Chamberland, Christopher and others},
  journal={Nat Commun.},
  volume={16},
  number={1},
  pages={10829},
  year={2025},
  publisher={Nature Publishing Group UK London}
}

@article{chuang1998experimental,
  title={Experimental realization of a quantum algorithm},
  author={Chuang, Isaac L and Vandersypen, Lieven MK and Zhou, Xinlan and Leung, Debbie W and Lloyd, Seth},
  journal={Nature},
  volume={393},
  number={6681},
  pages={143--146},
  year={1998},
  publisher={Nature Publishing Group UK London}
}

@article{jones1998implementation_seacrh,
  title={Implementation of a quantum search algorithm on a quantum computer},
  author={Jones, Jonathan A and Mosca, Michele and Hansen, Rasmus H},
  journal={Nature},
  volume={393},
  number={6683},
  pages={344--346},
  year={1998},
  publisher={Nature Publishing Group UK London}
}

@article{o2019hamiltonian,
  title={Hamiltonian engineering with constrained optimization for quantum sensing and control},
  author={O’Keeffe, Michael F and Horesh, Lior and Barry, John F and Braje, Danielle A and Chuang, Isaac L},
  journal={New J. Phys.},
  volume={21},
  number={2},
  pages={023015},
  year={2019},
  publisher={IOP Publishing}
}

@article{choi2020robust,
  title={Robust dynamic {H}amiltonian engineering of many-body spin systems},
  author={Choi, Joonhee and Zhou, Hengyun and Knowles, Helena S and Landig, Renate and Choi, Soonwon and Lukin, Mikhail D},
  journal={Phys. Rev. X},
  volume={10},
  number={3},
  pages={031002},
  year={2020},
  publisher={APS}
}

@article{zhou2024robust,
  title={Robust {H}amiltonian engineering for interacting qudit systems},
  author={Zhou, Hengyun and Gao, Haoyang and Leitao, Nathaniel T and Makarova, Oksana and Cong, Iris and Douglas, Alexander M and Martin, Leigh S and Lukin, Mikhail D},
  journal={Phys. Rev. X},
  volume={14},
  number={3},
  pages={031017},
  year={2024},
  publisher={APS}
}

@article{geier2021floquet,
  title={Floquet {H}amiltonian engineering of an isolated many-body spin system},
  author={Geier, Sebastian and Thaicharoen, Nithiwadee and Hainaut, Clement and Franz, Titus and Salzinger, Andre and Tebben, Annika and Grimshandl, David and Z{\"u}rn, Gerhard and Weidem{\"u}ller, Matthias},
  journal={Science},
  volume={374},
  number={6571},
  pages={1149--1152},
  year={2021},
  publisher={American Association for the Advancement of Science}
}

@article{zhou2023robust,
  title={Robust higher-order {H}amiltonian engineering for quantum sensing with strongly interacting systems},
  author={Zhou, Hengyun and Martin, Leigh S and Tyler, Matthew and Makarova, Oksana and Leitao, Nathaniel and Park, Hongkun and Lukin, Mikhail D},
  journal={Phys. Rev. Lett.},
  volume={131},
  number={22},
  pages={220803},
  year={2023},
  publisher={APS}
}

@article{ajoy2013quantum,
  title={Quantum Simulation via Filtered Hamiltonian Engineering: {A}pplication to Perfect Quantum Transport in Spin Networks},
  author={Ajoy, Ashok and Cappellaro, Paola},
  journal={Phys. Rev. Lett.},
  volume={110},
  number={22},
  pages={220503},
  year={2013},
  publisher={APS}
}

@article{ajoy2012algorithmic,
  title={Algorithmic approach to simulate {H}amiltonian dynamics and an {NMR} simulation of quantum state transfer},
  author={Ajoy, Ashok and Rao, Rama Koteswara and Kumar, Anil and Rungta, Pranaw},
  journal={Phys. Rev. A},
  volume={85},
  number={3},
  pages={030303},
  year={2012},
  publisher={APS}
}

@article{cory1997ensemble,
  title={Ensemble quantum computing by {NMR} spectroscopy},
  author={Cory, David G and Fahmy, Amr F and Havel, Timothy F},
  journal={Proc. Natl. Acad. Sci.},
  volume={94},
  number={5},
  pages={1634--1639},
  year={1997},
  publisher={The National Academy of Sciences of the USA}
}

@article{marx2000approaching,
  title={Approaching five-bit {NMR} quantum computing},
  author={Marx, R and Fahmy, AF and Myers, John M and Bermel, W and Glaser, SJ},
  journal={Phys. Rev. A},
  volume={62},
  number={1},
  pages={012310},
  year={2000},
  publisher={APS}
}

@article{anwar2004preparing,
  title={Preparing high purity initial states for nuclear magnetic resonance quantum computing},
  author={Anwar, M Sabieh and Blazina, Damir and Carteret, Hilary A and Duckett, Simon B and Halstead, TK and Jones, JA and Kozak, CM and Taylor, RJK},
  journal={Phys. Review Lett.},
  volume={93},
  number={4},
  pages={040501},
  year={2004},
  publisher={APS} 
}

@article{vandersypen2004nmr,
  title={{NMR} techniques for quantum control and computation},
  author={Vandersypen, Lieven MK and Chuang, Isaac L},
  journal={Rev. Mod. Phys.},
  volume={76},
  number={4},
  pages={1037--1069},
  year={2004},
  publisher={APS}
}

@article{alway2007arbitrary,
  title={Arbitrary precision composite pulses for {NMR} quantum computing},
  author={Alway, William G and Jones, Jonathan A},
  journal={J. Magn. Reson.},
  volume={189},
  number={1},
  pages={114--120},
  year={2007},
  publisher={Elsevier}
}

@article{peng2001preparation,
  title={Preparation of pseudo-pure states by line-selective pulses in nuclear magnetic resonance},
  author={Peng, Xinhua and Zhu, Xiwen and Fang, Ximing and Feng, Mang and Gao, Kelin and Yang, Xiaodong and Liu, Maili},
  journal={Chem. Phys. Lett.},
  volume={340},
  number={5-6},
  pages={509--516},
  year={2001},
  publisher={Elsevier}
}

@article{sharf2000spatially,
  title={Spatially encoded pseudopure states for NMR quantum-information processing},
  author={Sharf, Yehuda and Havel, Timothy F and Cory, David G},
  journal={Phys. Rev. A},
  volume={62},
  number={5},
  pages={052314},
  year={2000},
  publisher={APS}
}

@article{Plaumann2013,
author = {Plaumann, Markus and Bommerich, Ute and Trantzschel, Thomas and Lego, Denise and Dillenberger, Sonja and Sauer, Grit and Bargon, Joachim and Buntkowsky, Gerd and Bernarding, Johannes},
title = {Parahydrogen-Induced Polarization Transfer to $^{19}${F} in Perfluorocarbons for $^{19}${F} {NMR} Spectroscopy and {MRI}},
journal = {Chem. Eur. J.},
volume = {19},
number = {20},
pages = {6334-6339},
doi = {https://doi.org/10.1002/chem.201203455},
url = {https://chemistry-europe.onlinelibrary.wiley.com/doi/abs/10.1002/chem.201203455},
eprint = {https://chemistry-europe.onlinelibrary.wiley.com/doi/pdf/10.1002/chem.201203455},
year = {2013}
}

@article{SORENSEN1990435,
title = {A universal bound on spin dynamics},
journal = {J. Magn. Reson.},
volume = {86},
number = {2},
pages = {435-440},
year = {1990},
issn = {0022-2364},
doi = {https://doi.org/10.1016/0022-2364(90)90278-H},
url = {https://www.sciencedirect.com/science/article/pii/002223649090278H},
author = {Ole W S{\o}rensen}
}

@article{Kiryutin2024Photo,
author = {Kiryutin, Alexey S. and Kozinenko, Vitaly P. and Yurkovskaya, Alexandra V.},
title = {Photo-{SABRE}: Nuclear Spin Hyperpolarization of cis-trans Photoswitchable Molecules by Parahydrogen},
journal = {ChemPhotoChem},
volume = {8},
number = {1},
pages = {e202300151},
doi = {https://doi.org/10.1002/cptc.202300151},
url = {https://chemistry-europe.onlinelibrary.wiley.com/doi/abs/10.1002/cptc.202300151},
eprint = {https://chemistry-europe.onlinelibrary.wiley.com/doi/pdf/10.1002/cptc.202300151},
year = {2024}
}

@article{pravia1999observations,
  title={Observations of quantum dynamics by solution-state {NMR} spectroscopy},
  author={Pravia, Marco and Fortunato, Evan and Weinstein, Yaakov and Price, Mark D and Teklemariam, Grum and Nelson, Richard J and Sharf, Yehuda and Somaroo, Shyamal and Tseng, CH and Havel, Timothy F and others},
  journal={Concept. Magn. Reson. A},
  volume={11},
  number={4},
  pages={225--238},
  year={1999},
  publisher={Wiley Online Library}
}

@misc{kiryutin2026high,
title = {High-field NMR characterization and indirect J-spectroscopy of a nuclear spin chain [U-13C, 15N]-butyronitrile},
journal = {J. Magn. Reson. Open},
volume = {27},
pages = {100222},
year = {2026},
issn = {2666-4410},
doi = {https://doi.org/10.1016/j.jmro.2026.100222},
author = {Alexey Kiryutin and Ivan Zhukov and Danil Markelov and Erik {Van Dyke} and Alexandra Yurkovskaya and Danila Barskiy}
}

@article{du2010nmr,
  title={{NMR} Implementation of a Molecular Hydrogen Quantum Simulation with Adiabatic State Preparation},
  author={Du, Jiangfeng and Xu, Nanyang and Peng, Xinhua and Wang, Pengfei and Wu, Sanfeng and Lu, Dawei},
  journal={Phys. Rev. Lett.},
  volume={104},
  number={3},
  pages={030502},
  year={2010},
  publisher={APS}
}

@article{portmann2022security,
  title={Security in quantum cryptography},
  author={Portmann, Christopher and Renner, Renato},
  journal={Rev. Mod. Phys.},
  volume={94},
  number={2},
  pages={025008},
  year={2022},
  publisher={APS}
}

@article{bennett1992experimental,
  title={Experimental quantum cryptography},
  author={Bennett, Charles H and Bessette, Fran{\c{c}}ois and Brassard, Gilles and Salvail, Louis and Smolin, John},
  journal={J. Cryptol.},
  volume={5},
  number={1},
  pages={3--28},
  year={1992},
  publisher={Springer}
}

@article{deppner2021collectivemode,
  title   = {Collective-Mode Enhanced Matter-Wave Optics},
  author  = {Deppner, Christian and Herr, Waldemar and Cornelius, Merle and Stromberger, Peter and Sternke, Tammo and Grzeschik, Christoph and Grote, Alexander and Rudolph, Jan and Herrmann, Sven and Krutzik, Markus and Wenzlawski, Andre and Corgier, Robin and Charron, Eric and Guery-Odelin, David and Gaaloul, Naceur and L{\"a}mmerzahl, Claus and Peters, Achim and Windpassinger, Patrick and Rasel, Ernst M.},
  journal = {Phys. Rev. Lett.},
  volume  = {127},
  number  = {10},
  pages   = {100401},
  year    = {2021},
  doi     = {10.1103/PhysRevLett.127.100401}
}

@article{Yeter2021,
author = {Yeter-Aydeniz, Kübra and Gard, Bryan T. and Jakowski, Jacek and Majumder, Swarnadeep and Barron, George S. and Siopsis, George and Humble, Travis S. and Pooser, Raphael C.},
title = {Benchmarking Quantum Chemistry Computations with Variational, Imaginary Time Evolution, and Krylov Space Solver Algorithms},
journal = {Adv. Quant. Technol.},
volume = {4},
number = {7},
pages = {2100012},
doi = {https://doi.org/10.1002/qute.202100012},
url = {https://advanced.onlinelibrary.wiley.com/doi/abs/10.1002/qute.202100012},
eprint = {https://advanced.onlinelibrary.wiley.com/doi/pdf/10.1002/qute.202100012},
year = {2021}
}

@article{bowers1987parahydrogen,
  title={Parahydrogen and synthesis allow dramatically enhanced nuclear alignment},
  author={Bowers, C Russell and Weitekamp, Daniel P},
  journal={J. Am. Chem. Soc.},
  volume={109},
  number={18},
  pages={5541--5542},
  year={1987},
  publisher={ACS Publications}
}

@article{pravica1988net,
  title={Net {NMR} alignment by adiabatic transport of parahydrogen addition products to high magnetic field},
  author={Pravica, Michael G and Weitekamp, Daniel P},
  journal={Chem. Phys. Lett.},
  volume={145},
  number={4},
  pages={255--258},
  year={1988},
  publisher={Elsevier}
}

@article{ueda2023non,
  title={Non-equilibrium dynamics of spin-lattice coupling},
  author={Ueda, Hiroki and Mankowsky, Roman and Paris, Eugenio and Sander, Mathias and Deng, Yunpei and Liu, Biaolong and Leroy, Ludmila and Nag, Abhishek and Skoropata, Elizabeth and Wang, Chennan and others},
  journal={Nat. Commun.},
  volume={14},
  number={1},
  pages={7778},
  year={2023},
  publisher={Nature Publishing Group UK London}
}

@article{ALBERT2017293,
title = {Detecting multi-spin interactions in the inverse Ising problem},
journal = {Physica A Stat. Mech. Appl.},
volume = {483},
pages = {293-298},
year = {2017},
issn = {0378-4371},
doi = {https://doi.org/10.1016/j.physa.2017.04.120},
url = {https://www.sciencedirect.com/science/article/pii/S0378437117304521},
author = {Joseph Albert and Robert H. Swendsen}
}

@article{CADARS200724,
title = {The refocused {INADEQUATE} {MAS} {NMR} experiment in multiple spin-systems: Interpreting observed correlation peaks and optimising lineshapes},
journal = {J. Magn. Reson.},
volume = {188},
number = {1},
pages = {24-34},
year = {2007},
issn = {1090-7807},
doi = {https://doi.org/10.1016/j.jmr.2007.05.016},
url = {https://www.sciencedirect.com/science/article/pii/S1090780707001589},
author = {Sylvian Cadars and Julien Sein and Luminita Duma and Anne Lesage and Tran N. Pham and Jay H. Baltisberger and Steven P. Brown and Lyndon Emsley}
}

@article{lin2021real,
  title={Real-and imaginary-time evolution with compressed quantum circuits},
  author={Lin, Sheng-Hsuan and Dilip, Rohit and Green, Andrew G and Smith, Adam and Pollmann, Frank},
  journal={PRX Quantum},
  volume={2},
  number={1},
  pages={010342},
  year={2021},
  publisher={APS}
}

@article{schwade2026physics,
  title={Physics-informed Hamiltonian learning for large-scale optoelectronic property prediction},
  author={Schwade, Martin and Zhang, Shaoming and Vonhoff, Frederik and Delgado, Frederico P and Egger, David A},
  journal={Nat. Commun.},
  volume={2652},
  year={2026},
  publisher={Nature Publishing Group UK London}
}

@article{levitt2021hyperpolarization,
  title   = {Hyperpolarization and the physical boundary of Liouville space},
  author  = {Levitt, Malcolm H. and Bengs, Christian},
  journal = {Magn. Reson.},
  volume  = {2},
  pages   = {395--407},
  year    = {2021},
  doi     = {10.5194/mr-2-395-2021}
}

@misc{kozinenko2026,
      title={Improved {SABRE} hyperpolarisation using pulse sequences to reduce effective coupling}, 
      author={Vitaly P. Kozinenko and Bogdan A. Rodin and James Eills and Ilai Schwartz and Stephan Knecht and Laurynas Dagys},
      year={2026},
      eprint={2603.08622},
      archivePrefix={arXiv},
      primaryClass={physics.chem-ph},
      url={https://arxiv.org/abs/2603.08622}, 
}

@article{daley2022practical,
  title={Practical quantum advantage in quantum simulation},
  author={Daley, Andrew J and Bloch, Immanuel and Kokail, Christian and Flannigan, Stuart and Pearson, Natalie and Troyer, Matthias and Zoller, Peter},
  journal={Nature},
  volume={607},
  number={7920},
  pages={667--676},
  year={2022},
  publisher={Nature Publishing Group UK London}
}

@article{gershenfeld1997bulk,
  title={Bulk spin-resonance quantum computation},
  author={Gershenfeld, Neil A and Chuang, Isaac L},
  journal={Science},
  volume={275},
  number={5298},
  pages={350--356},
  year={1997},
  publisher={American Association for the Advancement of Science}
}

@article{sorensen1984product,
  title={Product operator formalism for the description of {NMR} pulse experiments},
  author={Soerensen, OW and Eich, GW and Levitt, Malcolm H and Bodenhausen, G and Ernst, RR},
  journal={Prog. Nucl. Magn. Reson. Spectrosc.},
  volume={16},
  pages={163--192},
  year={1984},
  publisher={Elsevier}
}

@article{sheberstov2025aliphatic,
  title={Aliphatic Chains as One-Dimensional {XY} Spin Chains},
  author={Sheberstov, Kirill F},
  journal={arXiv preprint},
  volume={2512.23759},
  year={2025}
}

@article{sheberstov2024collective,
  title={Collective long-lived zero-quantum coherences in aliphatic chains},
  author={Sheberstov, Kirill F and Sonnefeld, Anna and Bodenhausen, Geoffrey},
  journal={J. Chem. Phys.},
  volume={160},
  number={14},
  year={2024},
  publisher={AIP Publishing}
}

@article{sonnefeld2022polychromatic,
  title={Polychromatic excitation of delocalized long-lived proton spin states in aliphatic chains},
  author={Sonnefeld, Anna and Bodenhausen, Geoffrey and Sheberstov, Kirill},
  journal={Phys. Rev. Lett.},
  volume={129},
  number={18},
  pages={183203},
  year={2022},
  publisher={APS}
}

@article{sonnefeld2022long,
  title={Long-lived states of methylene protons in achiral molecules},
  author={Sonnefeld, Anna and Razanahoera, Aiky and Pelupessy, Philippe and Bodenhausen, Geoffrey and Sheberstov, Kirill},
  journal={Sci. Adv.},
  volume={8},
  number={48},
  pages={eade2113},
  year={2022},
  publisher={American Association for the Advancement of Science}
}

@article{ardenkjaer2003increase,
  title={Increase in signal-to-noise ratio of> 10,000 times in liquid-state NMR},
  author={Ardenkj{\ae}r-Larsen, Jan H and Fridlund, Bj{\"o}rn and Gram, Andreas and Hansson, Georg and Hansson, Lennart and Lerche, Mathilde H and Servin, Rolf and Thaning, Mikkel and Golman, Klaes},
  journal={Proc. Natl. Acad. Sci.},
  volume={100},
  number={18},
  pages={10158--10163},
  year={2003},
  publisher={National Academy of Sciences}
}

@article{berry2025rapid,
  title={Rapid initial-state preparation for the quantum simulation of strongly correlated molecules},
  author={Berry, Dominic W and Tong, Yu and Khattar, Tanuj and White, Alec and Kim, Tae In and Low, Guang Hao and Boixo, Sergio and Ding, Zhiyan and Lin, Lin and Lee, Seunghoon and others},
  journal={PRX Quantum},
  volume={6},
  number={2},
  pages={020327},
  year={2025},
  publisher={APS}
}

@article{anwar2004implementing,
  title={Implementing {G}rover’s quantum search on a para-hydrogen based pure state {NMR} quantum computer},
  author={Anwar, MS and Blazina, D and Carteret, HA and Duckett, SB and Jones, JA},
  journal={Chem. Phys. Lett.},
  volume={400},
  number={1-3},
  pages={94--97},
  year={2004},
  publisher={Elsevier}
}

@article{blazina2005generation,
  title={Generation and interrogation of a pure nuclear spin state by parahydrogen-enhanced {NMR} spectroscopy: a defined initial state for quantum computation},
  author={Blazina, Damir and Duckett, Simon B and Halstead, TK and Kozak, CM and Taylor, RJK and Anwar, M Sabieh and Jones, Jonathan A and Carteret, Hilary A},
  journal={Magn. Reson. Chem.},
  volume={43},
  number={3},
  pages={200--208},
  year={2005},
  publisher={Wiley Online Library}
}

@article{hubler2000nuclear,
  title={Nuclear magnetic resonance quantum computing exploiting the pure spin state of para hydrogen},
  author={H{\"u}bler, Patrick and Bargon, Joachim and Glaser, Steffen J},
  journal={J. Chem. Phys.},
  volume={113},
  number={6},
  pages={2056--2059},
  year={2000},
  publisher={American Institute of Physics}
}

@article{weber2022toward,
  title={Toward reliability in the {NISQ} era: Robust interval guarantee for quantum measurements on approximate states},
  author={Weber, Maurice and Anand, Abhinav and Cervera-Lierta, Alba and Kottmann, Jakob S and Kyaw, Thi Ha and Li, Bo and Aspuru-Guzik, Alan and Zhang, Ce and Zhao, Zhikuan},
  journal={Phys. Rev. Res},
  volume={4},
  number={3},
  pages={033217},
  year={2022},
  publisher={APS}
}

@article{rosenkranz2025quantum,
  title={Quantum state preparation for multivariate functions},
  author={Rosenkranz, Matthias and Brunner, Eric and Marin-Sanchez, Gabriel and Fitzpatrick, Nathan and Dilkes, Silas and Tang, Yao and Kikuchi, Yuta and Benedetti, Marcello},
  journal={Quantum},
  volume={9},
  pages={1703},
  year={2025},
  publisher={Verein zur F{\"o}rderung des Open Access Publizierens in den Quantenwissenschaften}
}

@article{gong2024complexity,
  title={Complexity of digital quantum simulation in the low-energy subspace: {A}pplications and a lower bound},
  author={Gong, Weiyuan and Zhou, Shuo and Li, Tongyang},
  journal={Quantum},
  volume={8},
  pages={1409},
  year={2024},
  publisher={Verein zur F{\"o}rderung des Open Access Publizierens in den Quantenwissenschaften}
}

@article{adams2009reversible,
  title={Reversible interactions with para-hydrogen enhance {NMR} sensitivity by polarization transfer},
  author={Adams, Ralph W and Aguilar, Juan A and Atkinson, Kevin D and Cowley, Michael J and Elliott, Paul IP and Duckett, Simon B and Green, Gary GR and Khazal, Iman G and Lopez-Serrano, Joaquin and Williamson, David C},
  journal={Science},
  volume={323},
  number={5922},
  pages={1708--1711},
  year={2009},
  publisher={American Association for the Advancement of Science}
}

@article{rayner2018signal,
  title={Signal amplification by reversible exchange ({SABRE}): From discovery to diagnosis},
  author={Rayner, Peter J and Duckett, Simon B},
  journal={Angew. Chem. Int. Ed.},
  volume={57},
  number={23},
  pages={6742--6753},
  year={2018},
  publisher={Wiley Online Library}
}

@article{lindale2024,
  title={Multi-axis fields boost SABRE hyperpolarization},
  author={Lindale, Jacob R and Smith, Loren L and Mammen, Mathew W and Eriksson, Shannon L and Everhart, Lucas M and Warren, Warren S},
  journal={Proc. Nat. Acad. Sci.},
  volume={121},
  number={14},
  pages={e2400066121},
  year={2024},
  publisher={National Academy of Sciences}
}

@article{eriksson2022,
  title={Improving SABRE hyperpolarization with highly nonintuitive pulse sequences: Moving beyond avoided crossings to describe dynamics},
  author={Eriksson, Shannon L and Lindale, Jacob R and Li, Xiaoqing and Warren, Warren S},
  journal={Sci. Adv.},
  volume={8},
  number={11},
  pages={eabl3708},
  year={2022},
  publisher={American Association for the Advancement of Science}
}

@article{truong201515n,
  title={$^{15}${N} hyperpolarization by reversible exchange using {SABRE-SHEATH}},
  author={Truong, Milton L and Theis, Thomas and Coffey, Aaron M and Shchepin, Roman V and Waddell, Kevin W and Shi, Fan and Goodson, Boyd M and Warren, Warren S and Chekmenev, Eduard Y},
  journal={J. Phys. Chem. C},
  volume={119},
  number={16},
  pages={8786--8797},
  year={2015},
  publisher={ACS Publications}
}

@article{theis2015microtesla,
  title={Microtesla {SABRE} enables 10\% nitrogen-15 nuclear spin polarization},
  author={Theis, Thomas and Truong, Milton L and Coffey, Aaron M and Shchepin, Roman V and Waddell, Kevin W and Shi, Fan and Goodson, Boyd M and Warren, Warren S and Chekmenev, Eduard Y},
  journal={J. Am. Chem. Soc.},
  volume={137},
  number={4},
  pages={1404--1407},
  year={2015},
  publisher={ACS Publications}
}

@article{chuang1998bulk,
  title={Bulk quantum computation with nuclear magnetic resonance: theory and experiment},
  author={Chuang, Isaac L and Gershenfeld, Neil and Kubinec, Mark G and Leung, Debbie W},
  journal={Proceedings of the Royal Society of London. Series A: Mathematical, Physical and Engineering Sciences},
  volume={454},
  number={1969},
  pages={447--467},
  year={1998},
  publisher={The Royal Society}
}

@article{kiryutin2013evidence,
  title={Evidence for coherent transfer of para-hydrogen-induced polarization at low magnetic fields},
  author={Kiryutin, Alexey S and Yurkovskaya, Alexandra V and Kaptein, Robert and Vieth, Hans-Martin and Ivanov, Konstantin L},
  journal={J. Phys. Chem. Lett.},
  volume={4},
  number={15},
  pages={2514--2519},
  year={2013},
  publisher={ACS Publications}
}

\end{document}